# The logistic equation and a critique of the theory of natural selection[1]

D. Balciunas

**Abstract**

Species coexistence is one of the central themes in modern ecology. Coexistence is a prerequisite of biological diversity. However, the question arises how biodiversity can be reconciled with the statement of competition theory, which asserts that competing species cannot coexist. To solve this problem natural selection theory is rejected because it contradicts kinetic models of interacting populations. Biological evolution is presented as a process equivalent to a chemical reaction. The main point is that interactions occur between self-replicating units. Under these assumptions biodiversity is possible if and only if species are identical with respect to the patterns of energy flow in which individuals are involved.

**I. The paradigm**

Competition theory, probably the oldest ecological theory (Gause, 1934b; Lotka), makes a fundamental link between ecology and evolutionary biology. The idea that competing individuals are evolving individuals, proposed by C. Darwin in the form of natural selection theory (Darwin, 1859), makes the background of contemporary evolutionary thought. Now this is widely accepted scientific model that explains the mechanism and describes the consequences of evolution of living things (Calow, 1983; Khare & Shaulsky, 2006).

According to Darwin the more similar species are the more severely they compete. This should lead to the extinction of some less adapted intermediate forms and to the further divergence of competing species. I name this statement on the relationship between species similarity and the strength of competitive interactions between them Darwin's competition principle (DCP). As we will see later it has close connections with other biological rule – the central law of ecology[2] – which has been called the principle of competitive exclusion (CEP) (Gilbert *et al.*, 1952; Hardin, 1960; Schoener, 1982). CEP, also called Gause's hypothesis, states the following: If we replace some individuals of a particular species, living in a given environment, with individuals of another species then these two populations cannot live together permanently – one species will be excluded, partially or completely, from that environment.

In the formulation of classical CEP and its more recent version – the limiting similarity theory (McArthur, 1967; Brown, 1981; Giller, 1984) – a considerable part took the Lotka-Volterra type (LV) species competition models (Gause; Lotka, 1925; Volterra, 1926; Pianka, 1978; van der Vaart, 1983). These equations have had a big impact on ecological theory and up to now are widely used (…).

The LV models are a natural extension of the logistic equation, which in turn is a modification of an exponential growth formula called the Malthus' growth law[3]. The following four equations show these models and relations between them.

1) The exponential growth function

$$dP/dt = rP \qquad (I.1)$$

2) The logistic equation

$$dP/dt = rP(1 - P/K) \qquad (I.2)$$

---

[1] This manuscript is an extended and revised version of my previous notes (Balciunas, 2004, 2005).
[2] More correctly we should declare CEP as an axiom (Løvtrup, 1986).
[3] This is not quite true as the essence of the Malthus' theory is, namely, the limited growth of a population (Malthus, 1798). Such an opinion was expressed by Richardson (1977).

3) The two-species LV competition model

$$dP_1/dt = r_1 P_1 (1 - P_1/K_1 - \alpha_{12} P_2/K_1)$$

$$dP_2/dt = r_2 P_2 (1 - P_2/K_2 - \alpha_{21} P_1/K_2) \qquad (I.3)$$

4) The system of LV competition equations describing *n*-species interactions

$$dP_i/dt = r_i P_i (1 - \alpha_{i1} P_1/K_i - \ldots - P_i/K_i - \ldots - \alpha_{in} P_n/K_i) \qquad (i = 1, \ldots, n) \qquad (I.4)$$

Here $P_i$ is the density of *i*th population individuals (or mass); $r_i$ – the specific growth rate parameter. $\alpha_{ik}$, an interspecific competition coefficient, shows the relative impact of one $P_k$ individual on the growth rate of $P_i$ population. All intraspecific competition coefficients, $\alpha_{kk}$, are equal to 1. $K_i$ is called the carrying capacity. It is the maximum number of individuals that a given environment may sustain. $K_i$ for every species $P_i$ in (I.2) and (I.3) is equal to K in the logistic equation for the same species, i.e when it grows without competitors. All these four equations may be considered as the logistic type equations. The exponential growth model is the logistic equation with $K \to \infty$.

Let us begin with a system of *n* competing species, which evolve according to the LV model (I.4)

$$dP_i/dt = r_i P_i (1 - \alpha_{i1} P_1/K_i - \ldots - P_i/K_i - \ldots - \alpha_{in} P_n/K_i) \qquad (i = 1, \ldots, n) \qquad (I.4)$$

This system reaches equilibrium when the following relations hold

$$\alpha_{i1} P_1/K_i + \ldots + P_i/K_i + \ldots + \alpha_{in} P_n/K_i = 1 \qquad (i = 1, \ldots, n) \qquad (I.5)$$

On the other hand, it follows from the LV model analysis that while

$$P_1/K_1 + \ldots + P_i/K_i + \ldots + P_n/K_n < 1 \qquad (I.6)$$

competing species cannot attain a stable state. Thus at equilibrium

$$P_1/K_1 + \ldots + P_i/K_i + \ldots + P_n/K_n \geq 1 \qquad (I.7)$$

and

$$1 - P_1/K_1 - \ldots - P_i/K_i - \ldots - P_n/K_n \leq 0 \qquad (I.8)$$

From here we may write

$$1 - P_i/K_i \leq P_k/K_k + \ldots + P_{i-1}/K_{i-1} + P_{i+1}/K_{i+1} + \ldots + P_n/K_n \qquad (i = 1, \ldots, n) \qquad (I.9)$$

Rewrite equations (I.5) in the following manner

$$1 - P_i/K_i - \alpha_{ik} P_1/K_i - \ldots - \alpha_{in} P_n/K_i = 0 \qquad (i = 1, \ldots, n) \qquad (I.10)$$

Replacing $1 - P_i/K_i$ in relation (I.10) with the right side expression of equation (I.9) we obtain

$P_1/K_1 + \ldots + P_{i-1}/K_{i-1} + P_{i+1}/K_{i+1} + \ldots + P_n/K_n -$

$\quad -\alpha_{i1}P_1/K_i - \ldots - \alpha_{i,i-1}P_{i-1}/K_i - \alpha_{i,i+1}P_{i+1}/K_i - \ldots - \alpha_{in}P_n/K_n \geq 0 \qquad (i = 1, \ldots, n) \quad (I.11)$

As $(\alpha_{kk}K_k - K_k) = 0$ for all $k$ we may write instead of (I.11)

$K_i^{-1}K_1^{-1}(\alpha_{i1}K_1 - K_i)P_1 + \ldots + K_i^{-1}K_n^{-1}(\alpha_{in}K_n - K_i)P_n \leq 0 \qquad (i = 1, \ldots, n) \qquad (1.12)$

Now we transform the equations (I.4) in such a way that the (I.12) expressions would be included into them

$dP_i/dt = r_iP_i(1 - P_1/K_1 - \ldots - P_n/K_n -$

$- K_i^{-1}K_1^{-1}(\alpha_{i1}K_1 - K_i)P_1 - \ldots - K_i^{-1}K_n^{-1}(\alpha_{in}K_n - K_i)P_n) \qquad (i = 1, \ldots, n)$

or more briefly

$dP_i/dt = r_iP_i(1 - \Sigma P_k/K_k - \Sigma K_i^{-1}K_k^{-1}(\alpha_{ik}K_k - K_i)P_k) \qquad (i = 1, \ldots, n) \qquad (I.13)$

The behavior of system (I.4) is quite complex (Strobeck, 1973). Of course $n$ species do not exclude each other if

$K_i^{-1}K_k^{-1}(\alpha_{ik}K_k - K_i)P_k < 0 \qquad (i, k = 1, \ldots, n) \qquad (I.14)$

It follows from (I.14) that

$\alpha_{ik}K_k - K_i < 0 \qquad (i, k = 1, \ldots, n) \qquad (I.15)$

So if we want to be assured that $n$ competitors will coexist permanently it is enough (Strobeck, 1973) to take all competition coefficients such that

$\alpha_{ik} < K_i/K_k \qquad (i, k = 1, \ldots, n) \qquad (I.16)$

I consider relations (I.16) as another way of presenting CEP. This may be true if relations (I.16) is derived from the two-species LV competition model (I.3) – they are the necessary conditions for the coexistence of two species.

Thus mathematics confirms that two competing species $P_1$ and $P_2$ according to the LV model may coexist permanently only if intraspecific competition is more intense than interspecific competition, i. e. if $1/K_i > \alpha_{ki}/K_k$, $(i, k = 1, 2)$, or when $\alpha_{ik}\alpha_{ki} < 1$. This fact ecologically has been interpreted as a shift of species niches (McArthur, 1967).

Now we look again at the equations (I.12) and suppose that all $n$ species interact in that way that

$\Sigma K_i^{-1}K_k^{-1}(\alpha_{ik}K_k - K_i)P_k = 0 \qquad (i = 1, \ldots, n) \qquad (I.17)$

Our $n$-species LV competition model thus become a set of equations

$dP_i/dt = r_iP_i(1 - \Sigma(P_k/K_k)) \qquad (i = 1, \ldots, n) \qquad (I.18)$

Every combination of species densities $P_i > 0$ $(i = 1, \ldots, n)$ which satisfy relation

$$\Sigma P_k/K_k = 1 \tag{I.19}$$

is a solution for this system. Because here, at equilibrium, as well $\Sigma \alpha_{ik} P_k/K_i = 1$ ($i = 1, \ldots, n$), we obtain from (I.17) the following necessary expressions for competition coefficients $\alpha_{ik}$

$$\alpha_{ik} = K_i/K_k \qquad (i, k = 1, \ldots, n) \tag{I.20}$$

Species, which interact according to the model (I.18), I will call identical. They cannot exclude each other. If in the original LV competition model (I.4) we take all $\alpha_{ik} = 1$ then species will coexist only when their carrying capacities $K_i$ will be equal.

This special case of species coexistence (the fifth outcome of two-species LV competition model)[4] has not attracted much attention from mathematicians or theoretical ecologists. However, it will play a crucial role further.

## II. Contradictions and controversies

In the previous chapter we saw that the condition of species coexistence, derived from the LV competition models, is in agreement with DCP – Darwin's view on how intensively differently related species compete. However, this is the case only if we look at a stable ecological community, i.e. such in which species coexist. If, as Darwin stated, some individuals compete more strongly with individuals of their own species than with individuals of potential species-competitors, then, according to the LV competition models, competitive exclusion will not occur. It is clear: the more closely related species we take the more strongly they would compete. But anyway intraspecific competition would remain the strongest. If such a situation is a rule then every new form of life that appear will survive and natural selection would not work. If we rely on mathematical models, for Darwinian evolution to proceed there should be a situation between interacting species when intraspecific competition is less intense than interspecific competition. We have a paradoxical situation. Darwin, trying to explain species evolution, prevented itself from doing that.

DCP describes a community of interacting species after all rearrangements in community structure are finished. That means it is an extreme case of CEP, the validity of was discussed in a number of papers. CEP has received both criticism (Ayala, 1960, 1971b; Cole, 1960; Savile, 1960; Turner, 1970; Hulley *et al.*, 1988; Walter, 1988) and support (Gause, 1934a, 1970; Hardin, 1960; Van Valen, 1960; Borowsky, 1971; Antonovics and Ford, 1972; Gilpin and Justice, 1972). The phenomenon of species coexistence due to the aggregation of rather should be considered as a rejection of CEP populations (Kuno, 1988; Britton, 1989; O'Connell, 1996; but see Green, 1986). Den Boer (1980, 1985, 1986) instead of CEP proposed the coexistence principle because, according to him, closely related species are more prone to live together than may be expected. However, opposite evidence have been presented by Maherali *et al.* (2007): they found that distantly related species are more likely to coexist. Views similar to those of Den Boer's expressed Bengtsson (1986), Azovsky (1992, 1996), Huisman and Weising (1999). Such observations and theoretical constructions may be attributed to the 'plankton paradox' (Ghilarov, 1984; Hobson, 1988/1989): How it may be that a lot of similar species coexist in the seemingly same niche?

On the other hand, limiting similarity theory, created by McArthur (1967), is able to explain how species can be involved in Darwinian natural selection process. However, the coexistence of similar species also remains a problem for limiting similarity theory. This variant of CEP states that species cannot exclude each other only below some similarity level. Here similarity is determined in the terms of the overlaps of species niches. If species are more similar than the critical value, then they cannot coexist permanently.

---

[4] Four standard outcomes for this system are stable coexistence, unstable coexistence, first species wins, second species wins (Pianka, 1978).

According to LV competition model those species which have the same strength of intraspecific and interspecific interactions would be similar. If we require that

$$1/K_k = \alpha_{ik}/K_i = \alpha_{ki}/K_k = 1/K_i \tag{II.1}$$

then it follows from here that

$$\alpha_{ik} = \alpha_{ki} = 1 \text{ and } K_i = K_k \tag{II.2}$$

Such species are identical, but they will not exclude each other. Thus, CEP will not be supported. So the case $K_i = K_k$ should be excluded from the definition of CEP. For example, Ayala (1971a) begins the description of CEP by the words: "The [CEP] postulates that no two species are likely to be exactly identical in their efficiency to exploit any given resource."

The idea lying behind CEP is presented schematically in Fig. 1a. To be similar means to have some common requirements. No doubts that Darwin had this in mind when he wrote that more similar species compete more strongly.

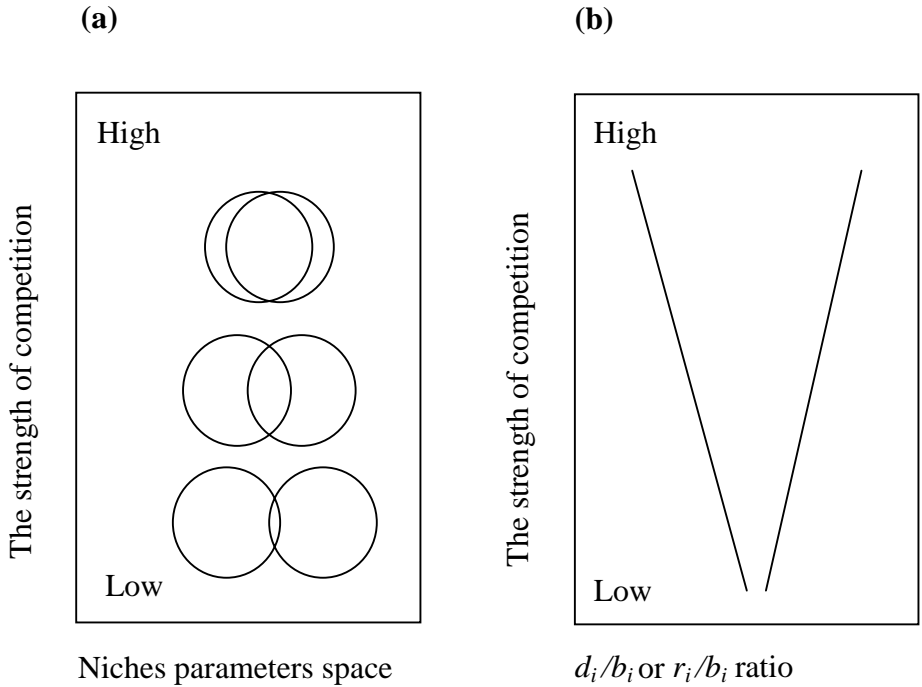

*Figure 1*. Two different approaches to competitive interactions. (a) Each circle represents a niche occupied by a single species. The differences between species are expressed in the units of niches overlap. Interspecific competition is a function of this overlap. The more close niches of two species are the more severely these species compete. This approach stresses only differences between niches but do not say anything how much similar individuals of both species are with respect to the use of resources. (b) The resource competition theory relates competition strength

with intrinsic individual characteristics – biochemical, physiological or behavioral. In the case of intraspecific exploitative competition the strength is evaluated by the efficiency of resource utilization. It is expressed as a ratio $d_i/b_i = R_i^*$, where $d_i$ and $b_i$ are the population mortality rate per capita and the population birth rate per capita on the unit amount of resource, respectively. Apparent competition gives analogous criteria $r_i/b_i = P_i^*$. Here $r_i$ is the resource growth rate and $b_i$ – the resources consumption rate. $R^*$ is a minimum amount of resources on which consumer can still coexist. $P^*$ is a maximum of a consumer amount which a given resource can support. The lines hypothetically show this ratio for two species. Similarity between species is calculated as the similarity between their $d_i/b_i$ (or $r_i/b_i$) ratios. This approach states that the more close species are the less they compete. The niches of both species in this case is assumed to be identical (individuals of both species exploit the same resource). Exploitation of the same niche is a necessary condition for competition to occur. So any primary concepts about competitive interactions should be based on the analysis of interactions between individuals occupying the same niche.

An opposite view on the similarity between species presents the resource competition theory (Hansen and Hubbell, 1980; Tilman, 1976). This approach looks at how species consume their common resources. If two or more species have the same pattern of resource consumption then these species are similar (Fig. 1b). Species of consumers are identical if they require the same minimal amount of resources $R_i^*$ for maintaining their populations. This criterion may also be used ( Holt *et al.*, 1994) in the case of apparent competition (Holt, 1977). Several resources are identical if they can feed the same maximum density $P_i^*$ of a consumer population.

The resource competition theory reformulates CEP in the following way: The number of coexisting species, competing for the same resources cannot exceed the number of resources. The same is true for apparent competition. In more general form this means that the number of coexisting species is no more than the number of controlling factors (Williamson, 1957). However, the possibility of coexistence of at least two identical species on a single resource (Tilman, 1981) undermines this variant of CEP too.

Take $\alpha_{ik} = 1$ ($i, k = 1, …, n$) in the LV competition model and assume that the carrying capacities $K_i$ is a function of $R_i^*$ parameter derived from resource competition models. Then LV competition model would turn to equations explicitly describing resource competition – the fact that the resource competition theory does not want to recognize (Tilman, 1987a, b; Grover, 1997). Then competitors identity in LV model will be related with species exploitation capabilities and will have nothing to deal with their niches overlapping. We come to the conclusion that the more similar species are the less they compete. Competitive exclusion can occur only if species have different $K_i$ values. This idea has led to the theory of limiting dissimilarity (Ågren and Fagerström, 1984; Aarssen, 1989).

But if identical species coexist then, maybe, all species, which coexist, are identical? If so, the problem of how biodiversity is maintained disappears. We see a huge number of species on earth simply because they all are identical and, thus, cannot exclude each other.

**III. Indifferent and directed competition**

Suppose that two species are grown together and consume the same food. We have an example of exploitative competition between two populations of consumers. No other interactions between the competitors exist. In this case the interactions between these two species may be modeled by a set of equations

$$dP_1/dt = r_1 P_1 (1 - P_1/K_1 - P_2/K_1)$$

$$dP_2/dt = r_2 P_2 (1 - P_2/K_2 - P_1/K_2) \qquad (III.1)$$

(III.1) may be supported by the following reasoning. At first we should assume that the mass of every individual of both species is equal or we should consider not a density of individuals but a density of their populations total mass. If a mass unit of the first species $P_1$ affects its own growth rate by a value $1/K_1$, then, it is clear, that the same mass unit of species $P_2$ will affect the growth rate of first species by the same value, i.e. equal to $1/K_1$. The same is true for the second species $P_2$. Its growth rate will be affected by the same value $1/K_2$ irrespective of interactions with what species, $P_1$ or $P_2$, we consider. Further in this chapter the term 'competition' will be used only for exploitative competitive interactions.

Now we will carry out the following experiment. Let us say that we have two species $P_1^l$ and $P_2^l$. They have originated from the parental species P by divergent evolution $P \rightarrow P_1^1 \rightarrow P_1^2 \rightarrow \ldots \rightarrow P_1^n$ and $P \rightarrow P_2^1 \rightarrow P_2^2 \rightarrow \ldots \rightarrow P_2^n$ (*Fig. 2*).

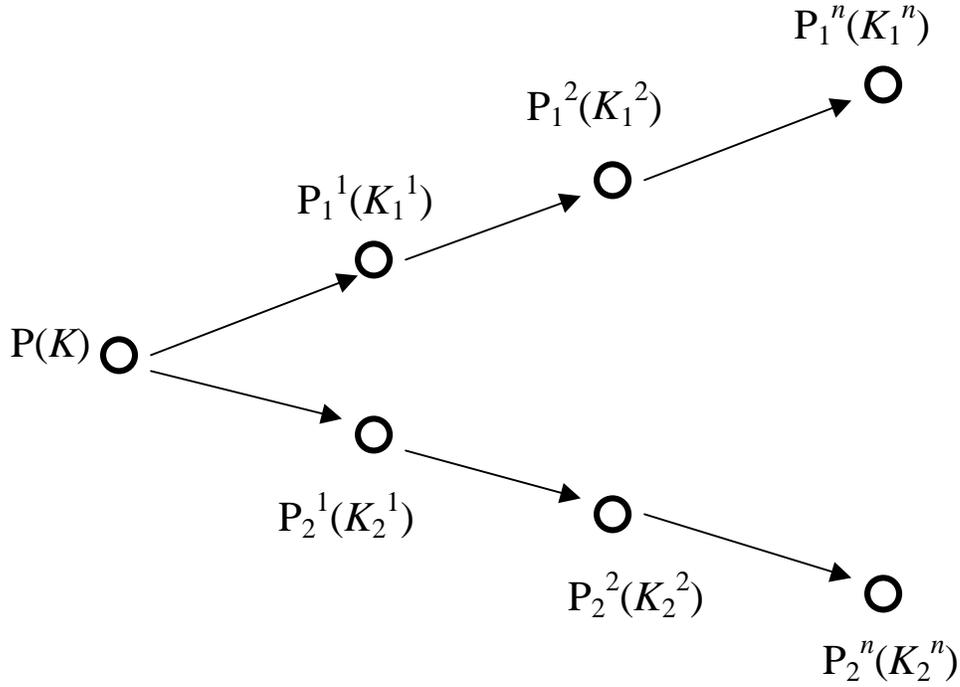

*Figure 2*. The relationship between species $P_1$ and $P_2$ involved in a hypothetical competition experiment. For all $l$ $K_1^l > K_2^l$, besides $K_1^r - K_2^r > K_1^s - K_2^s$ when $r > s$.

Both species $P_1^l$ and $P_2^l$ as well as initial species P use the same resource R. But species $P_1^l$ and $P_2^l$ cannot coexist stably on this resource. One species, for example $P_1^l$, will always exclude the other, in this case $P_2^l$. Both species have evolved in such a manner that the difference between their carrying capacities values $K_1^l$ and $K_2^l$ increases in the course of evolution. Our goal is to cultivate both species $P_1^l$ and $P_2^l$ on the same resource and to measure the strength of interspecific competition between them. This strength we may estimate by measuring the time interval needed for one species (in our case $P_1^l$) to exclude other ($P_2^l$). Growing species $P_1^l$ and $P_2^l$ together we will see that the closer species are to the original species P the less they compete, i.e. the more time is needed for the first species $P_1^l$ to exclude the second species $P_2^l$. This means that competition will disappear when species $P_1^l$ and $P_2^l$ converge to P. From here we may conclude that individuals of species P do not compete, i.e. that intraspecific competition do not exist. But we have to recognize that in fact intraspecific competition between individuals of species P exists. It is clear because we assumed that P species had carrying capacity *K*. How it might be that after disappearing of interspecific competition suddenly appears intraspecific competition? Probably the answer is such that intraspecific competition was always presented in our system of two competing species. What we earlier called interspecific competition is a summ of intraspecific competition and some other specific form of competitive interactions Δ. Only this specific form of competition disappears when species become more and more similar. This form of competition I called directed competition (Balciunas, 2004). We thus may write

Interspecific competition = intraspecific competition + Δ  (III.2)

Δ → 0 when $P_1^l$ → P and $P_2^l$ → P. When Δ = 0 species cannot be recognized by competitive interactions and they behave as identical species. Competition between identical individuals I called indifferent competition (Balciunas, 2004). From here competition (which in classical terms may be intraspecific as well as interspecific) is equal to

Competition = Indifferent competition + Directed competition  (III.3)

The notions of intraspecific and interspecific competition are quite inappropriate for theory. They do not reveal the essence of competitive interactions and thus cannot correctly describe the process of evolution of interacting species.
From a set of equations

$dP_1/dt = r_1 P_1 (1 - P_1/K_1 - \ldots - P_n/K_1 - K_1^{-1} K_1^{-1} (K_1 - K_1) P_1 - \ldots - K_1^{-1} K_n^{-1} (K_n - K_1) P_1)$
…
$dP_n/dt = r_n P_n (1 - P_1/K_n - \ldots - P_n/K_n - K_n^{-1} K_1^{-1} (K_1 - K_n) P_1 - \ldots - K_n^{-1} K_n^{-1} (K_n - K_n) P_1)$  (III.4)

we find the expressions of indifferent and directed competition. Indifferent competition is equal to a sum Σ ($P_k / K_k$). It is the same for all competing species. Contrary to indifferent competition directed competition in general may be different for different species. For a particular species $P_i$ it is a sum Σ ($K_1^{-1} K_n^{-1} (K_n - K_1) P_1$). The antisymmetric matrix Λ = ($\lambda_{ik}$) may be named the evolutionary matrix. Here $\lambda_{ik} = (K_i^{-1} K_k^{-1} (K_k - K_i))$, and $\lambda_{ik} = -\lambda_{ki}$. This matrix shows which species will be excluded from a system of interacting species.
Instead of CEP, which concentrates on exclusion the following principle, should apply to competing species. Species cannot coexist if there is directed competition between them. If species coexist they are identical.
As the outcome of species competition depends on the difference $K_k - K_i$ we might come to the idea that this expression, not natural selection, is a real force of biological evolution.

**IV. Self-replicating systems**

Consider a system with a total mass density M = M(*t*), here *t* means time. Suppose that the density does not change over time, i.e. M = *const*. The system is made of two components - X and Y so that

M = X + Y  (IV.1)

Component Y is self-replicating. It uses some amount of the component X to make a copy of itself, thus increasing its own mass density.

Suppose that after some time densities of both components, X and Y, reach steady states $X^*$, $Y^*$.

$$M = X^* + Y^* \tag{IV.2}$$

From (IV.1) and (IV.2) we obtain

$$X + Y = X^* + Y^* \tag{IV.3}$$

and further

$$X - X^* = Y^* - Y \;\Rightarrow\; X - X^* = Y^*(1 - Y/Y^*) \;\Rightarrow$$

$$\Rightarrow\; (X - X^*) / Y^* = 1 - Y/Y^* \tag{IV.4}$$

Now we make an obvious but important replacement. As $Y^* = M - X^*$, rewrite (IV.4) in the following form

$$(X - X^*) / (M - X^*) = 1 - Y/Y^* \tag{IV.5}$$

Equation (IV.5) is a basic expression.

Let us say that some function $V$ exists, that describes the rate by which the mass density of component Y changes over time, such that

$$V := f(X - X^*) \tag{IV.6}$$

Because component Y is self-replicating, the growth rate $V$ will approach zero value at equilibrium, i.e. when $Y = Y^*$, and also when $Y \to 0$ ($X \to M$). So we introduce a new function

$$v := V / Y = f(X - X^*) / Y = g(X - X^*) \tag{IV.7}$$

This last function also depends on $X - X^*$ as $Y = Y^* - (X - X^*)$. If function $g$ is single-valued we may write the inverse function of $g$

$$X - X^* = g^{-1}v \tag{IV.8}$$

Suppose that function $g$ has a limit value at the point $Y = 0$, and introduce the following notation

$$\varphi := g(M - X^*) \tag{IV.9}$$

It means that $v \to \varphi$, when $X \to M$ ($Y > 0$). So we have

$$M - X^* = g^{-1}\varphi \tag{IV.10}$$

Substituting the left side of equation (IV.5) with functions $v$ and $\varphi$ we obtain

$$g^{-1}v / g^{-1}\varphi = 1 - Y/Y^* \tag{IV.11}$$

or

$$g^{-1}v = (1 - Y/Y^*) g^{-1}\varphi \quad \text{(IV.12)}$$

Further we may write

$$gg^{-1}v = g((1 - Y/Y^*) g^{-1}\varphi) \quad \text{(IV.13)}$$

If we accept that function *g* is linear, equation (IV.13) will turn into

$$gg^{-1}v = (1 - Y/Y^*) g\, g^{-1}\varphi \quad \text{(IV.14)}$$

(IV.14) follows from the *Fig. 3*.

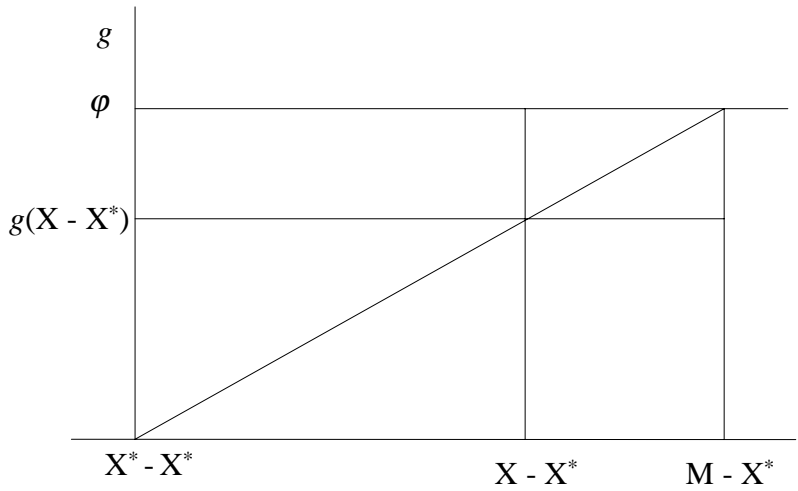

*Figure 3*. The equivalence of equations (IV.13) and (IV.14) when *g* is a linear function.
$(X - X^*) g (M - X^*) = (M - X^*) g (X - X^*) \Rightarrow$
$\Rightarrow ((X - X^*) / (M - X^*)) g (M - X^*) = g ((X - X^*)(M - X^*) / (M - X^*)) \Rightarrow$
$\Rightarrow ((X - X^*) / (M - X^*)) g (g^{-1}\varphi) = g (((X - X^*) / (M - X^*)) g^{-1}\varphi)$

From (IV.14) we finally get

$$v = (1 - Y/Y^*)\varphi \quad \text{(IV.15)}$$

As $v := V/Y$, we may further change formula (IV.15)

$$V/Y = (1 - Y/Y^*)\varphi \quad \text{(IV.16)}$$

And at last we have the final expression

$$V = \varphi Y (1 - Y/Y^*) \quad \text{(IV.17)}$$

If we denote $V := dY/dt$ and $Y^* := \Phi$, we obtain the rate equation for our system (IV.1)

$$dY/dt = \varphi Y (1 - Y/\Phi) \tag{IV.18}$$

Because function $g$ is a linear function we may write

$$V / Y = g (X - X^*) \tag{IV.7}$$

or

$$V = Yg (X - X^*) \tag{IV.19}$$

As our linear function $g$ has the expression

$$g (X - X^*) = b (X - X^*) \tag{IV.20}$$

where $b$ is some constant, we have

$$V = Yb (X - X^*) \tag{IV.21}$$

or

$$V = bXY - bX^*Y \tag{IV.22}$$

From (IV.22) we obtain

$$V = bXY - dY \tag{IV.23}$$

Here we use the following notation

$$d = bX^* \Rightarrow X^* = d/b \tag{IV.24}$$

The expression for $\Phi$ can be easily found from relation $Y^* = M - X^*$ and (IV.24)

$$\Phi = M - d/b = M - c;\ c = d/b \tag{IV.25a}$$

To find $\varphi$ let us write again (IV.21) and make the following rearrangements

$$V = Yg (X - X^*) \Rightarrow V = Yb (X - X^*) = Yb (M - Y - X^*) = bMY - bX^*Y - bY^2 \Rightarrow$$

$$\Rightarrow V = (bM - bX^*) Y - bY^2 = (bM - bX^*) Y (1 - Y/(M - X^*)) = \varphi Y (1 - Y/(M - X^*))\ \text{if}\ \varphi \neq 0.$$

Thus

$$\Phi = M - X^* = \varphi/b = (bM - bX^*) / b \tag{IV.25b}$$

where

$$\varphi = bM - bX^* = bM - d \tag{IV.26}$$

Parameters $b$ and $d$ have a meaning of rate constants of the forward (X → Y) and reversible (Y → X) processes respectively.

Model (IV.23) presents a formal chemical autocatalytic reaction – (self-replication) (Appendix 1). Such a reaction may be formally written in the form

X + Y → Y + Y

Y → X  (IV.27)

Let us begin with an initial rate equation describing the evolution of (IV.27) system

$dY/dt = bXY - dY$  (IV.28)

From here we obtain

$dY/dt = bXY - dY = b(M - Y)Y - dY = bMY - dY - bY^2 = (bM - d)Y(1 - Y/\Phi); bM - d \neq 0$

This is a usual procedure, which we will use to transform models of (IV.28) type to the (IV.18) expression. This model is not the logistic equation. Indeed, if we have $\Phi \to \infty$ then $\varphi$ also become infinite (because $b$ is finite). We do not obtain the exponential growth model if we suppose that $\Phi$ is infinitely large. But the exponential growth model can be written in (IV.18) form. Consider differential equation

$dY/dt = rY = (bX_0 - d)Y = bX_0Y - dY$

Because $r$ is constant it is equal to $bX_0 - d$. Thus

$dY/dt = (bX_0 - d)Y = bX_0Y - dY$

Here $X_0$ is some constant amount of resources, which we maintain at a given level by, for example, introducing a fresh amount of resources at the same rate at which they are converted to the product. Thus we constantly enhance the total mass density M of our system. We have further

$dY/dt = b(M - Y)Y - dY = (bM - d)Y(1 - Y/(M-d/b))$

As $M = X_0 + Y$ we may write

$dY/dt = (bM - d)Y(1 - Y/(Y + X_0 - d/b))$

If $d/b < X_0$ the concentration of Y will increase. Its $\Phi$ value will increase in the course of the system evolution too.

Regarding system (IV.27) we may write the expression analogous to (IV.18) for the component X also.

$dX/dt = -bXY + dY = b(M - X)X + dY = -bMX - dY - bX^2 =$

$= -(bM - dY/X)X(1 - Y/(M - dY/bX))$

or more briefly

$$dX/dt = -\varphi X(1 - Y/\Phi) \qquad (IV.29)$$

where $\varphi = (bM - dY/X)$ and $\Phi = M - dY/bX$.

Model (IV.27) is a closed variant of the simplest LV predator-prey system (Appendix 2).

$$dX/dt = rX - bXY$$

$$dY/dt = bXY - dY \qquad (IV.30)$$

These equations give the following expressions of type (IV.18)

$$dX/dt = -(bM - r)X(1 - X/(M - r/b)) \qquad (IV.31)$$

$$dY/dt = (bM - d)Y(1 - Y/(M - d/b)) \qquad (IV.32)$$

When $rX = dY$, i.e. $M = const$, from (IV.32) we obtain equation (IV.18) with $M - d/b = const$. When we move from open system (IV.30) to closed system (IV.27) oscillations which are characteristic for LV predator-prey model (IV.30) disappear (Appendix 3).

Now proceed to the next step. Suppose at first that we have n types of Y components or $m$ types of X components. It is more usual to work with different Y components. Thus we consider a system where $n$ $Y_i$ components self-replicate on the same resource X. The model describing interactions between the components is

$$dX/dt = \Sigma d_k Y_k - \Sigma b_k XY_k$$

$$dY_i/dt = b_i XY_i - d_i Y_i \qquad (i = 1, \ldots, n) \qquad (IV.33)$$

For every $Y_i$ type we obtain

$$dY_i/dt = b_i(M - Y_1 - \ldots - Y_n)Y_i - d_i Y_i = (b_i M - d_i)Y_i(1 - Y_1/(M - d_i/b_i) - \ldots - Y_n/(M - d_i/b_i)) =$$

$$= \varphi_i Y_i(1 - Y_1/\Phi_i - \ldots - Y_n/\Phi_i) \qquad (i = 1, \ldots, n) \qquad (IV.34a)$$

We obtained an expression, which we used earlier in the previous chapters. However, I should to stress that (IV.34) is not the LV competition model (See Appendix 4).

For $m$ competing resources $X_i$ we write according to (IV.29)

$$dX_i/dt = -\varphi_i X_i(1 - X_1/\Phi_i - \ldots - X_m/\Phi_i) \qquad (i = 1, \ldots, m) \qquad (IV.34b)$$

Models (IV.34a, b) describe virtual interactions between different forms of the same component (*Fig. 3*). I call these virtual interactions competition although the more appropriate term for them would be concurrency[5]. From now, when I will talk about competition, I will have in mind only these virtual interactions. Do not confuse them with what biologists call competition.

Equations (IV.34a) and (IV.34b) may be rewritten in one general form

$$dU_i/dt = \pm\varphi_i(U_i - U_i U_1/\Phi_i - \ldots - U_i U_n/\Phi_i) \qquad (i = 1, \ldots, l); \quad l \in \{m, n\} \qquad (IV.35)$$

---

[5] Such parallel chemical reactions are named concurrent reactions.

Here $U_i$ is either $X_i$ or $Y_i$ depending on our needs; $\Phi_i = M - c_{(i)}$.

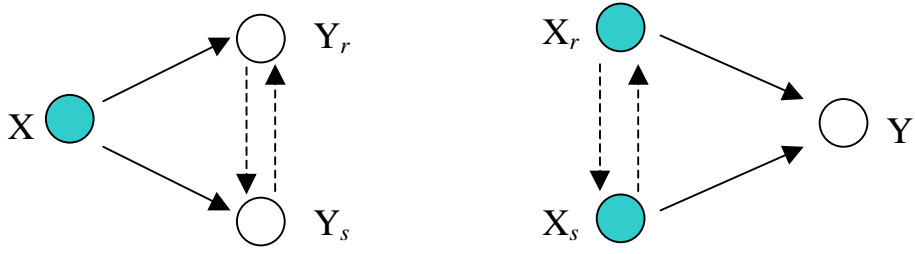

Divergent (d-) competition         Convergent (c-) competition

*Figure 3*. Virtual interactions between d- and c-competitors.

As we did earlier we now split competition into two parts – symmetrical interactions (indifferent competition) and antisymmetrical interactions (directed competition). Remember that $U_i$ is equivalent to $\Phi_i \Phi_k^{-1} U_k$. Thus symmetrical competition is presented as

$$(1/\Phi_i)U_i(\Phi_i/\Phi_k)U_k = (1/\Phi_i)(\Phi_i/\Phi_k)U_i U_k = (1/\Phi_k) U_i U_k \qquad (IV.36)$$

and antisymmetrical competition – as

$$\lambda_{ik} U_i U_k = (1/\Phi_i - 1/\Phi_k) U_i U_k = \Phi_i^{-1}\Phi_k^{-1} (\Phi_k - \Phi_i) U_i U_k = \Phi_i^{-1}\Phi_k^{-1} (c_{(i)} - c_{(k)}) U_i U_k \qquad (IV.37)$$

Therefore we have

$$dU_i/dt = \pm \varphi_i U_i (1 - \Sigma U_k/\Phi_k - \Sigma \lambda_{ik} U_k) \qquad (IV.38)$$

The matrix $\Lambda = (\lambda_{ik})$ is, as we described earlier, an antisymmetric matrix of directed competition functions.

We know that competitors may coexist only if

$$\Sigma_k(\lambda_{ik} U_k) = 0, \qquad (i, k = 1, \ldots, l); \quad l \in \{m, n\} \qquad (IV.39)$$

i.e. when $\lambda_{ik} = 0$, $(i, k = 1, \ldots, l)$, $l \in \{m, n\}$. That means that matrix $\Lambda = (0)$.

Now consider a more general case − a model describing trophic relations, and perhaps other types of interactions, in some hypothetical ecosystem

$$dU_i/dt = \pm(\Sigma \beta_{ik} U_i - \rho_i U_i) \qquad (i, k = 1, \ldots, l); \quad l \in \{m, n\} \qquad (IV.40)$$

Here $\beta_{ik}$ and $\rho_i$ are some functions. If total mass density of the ecosystem is $M(t)$ we may write

$$dU_i/dt = \pm(\Sigma \beta_{ik} W_k W_k^{-1} U_i - \rho_i U_i)$$

$$dU_i/dt = \pm(\Sigma \beta_{ik} W_k^{-1}(M - \Sigma_{r \neq k} W_r - \Sigma U_k - \Sigma S_p) U_i - \rho_i U_i) \qquad (IV.41)$$

where $W_k$ are resources for which $U_i$ compete or consumers which prey on $U_i$ resource; $S_p$ – species other than $W_k$ and $U_i$. Continuing from (IV.41)

$$dU_i/dt = \pm((\Sigma\beta_{ik}W_k^{-1}M - \Sigma\beta_{ik}W_k^{-1}\Sigma_{r \neq k}W_r - \Sigma\beta_{ik}W_k^{-1}\Sigma S_p - \rho_i)U_i - \Sigma\beta_{ik}W_k^{-1}\Sigma U_k U_i)$$

Denote $\varphi_i = \Sigma\beta_{ik}W_k^{-1}M - \Sigma\beta_{ik}W_k^{-1}\Sigma_{r \neq k}W_r - \Sigma\beta_{ik}W_k^{-1}\Sigma S_p - \rho_i$. If $\varphi_i \neq 0$ we write

$$dU_i/dt = \pm\varphi_i U_i(1 - \Sigma U_k \Phi_k^{-1}) \tag{IV.42}$$

Here $\Phi_k = \varphi_i/\Sigma\beta_{ik}W_k^{-1} = M - \Sigma S_p - (\Sigma\beta_{ik}W_k^{-1}\Sigma_{r \neq k}W_r)/\Sigma\beta_{ik}W_k^{-1} - \rho_i/\Sigma\beta_{ik}W_k^{-1}$.

As we do not consider exponential growth (the mass of a physical system is limited) then

$$\lim_{t \to \infty} t^{-1} \int_0^t (dU_i/d\tau)\, d\tau = 0 \tag{IV.43}$$

Because

$$\lim_{t \to \infty} t^{-1} \int_0^t \varphi_i U_i\, d\tau \neq 0 \tag{IV.44}$$

(this means that species $U_i$ does not die out) we have

$$\lim_{t \to \infty} t^{-1} \int_0^t (1 - \Sigma U_k/\Phi_k - \Sigma\lambda_{ik}U_k)d\tau = 0 \tag{IV.45}$$

As

$$\lim_{t \to \infty} t^{-1} \int_0^t (1 - \Sigma U_k/\Phi_k)d\tau = 0 \tag{IV.46}$$

then from here it follows that

$$\lim_{t \to \infty} t^{-1} \int_0^t \Sigma\lambda_{ik}U_k d\tau = 0 \tag{IV.47}$$

If we want that equation (IV.47) would be satisfied it is necessary that, as we saw earlier, all directed competition functions would have a mean value of zero, i.e.

$$\lim_{t \to \infty} t^{-1} \int_0^t \lambda_{ik}d\tau = 0, \qquad (i, k = 1, \ldots, l); \quad l \in \{m, n\} \tag{IV.48}$$

If species coexist they must be identical at least in average.

Now we return to the formal chemical reactions. Let us consider two of them – one parallel and one concurrent (*Fig. 4*). The parallel chemical reaction (*Fig. 4a*) is described by the scheme

$$\begin{array}{ll} X \to Y; & Y \to X \quad (\text{or } X \leftarrow\to Y) \\ X \to Z; & Z \to X \quad (\text{or } X \leftarrow\to Z) \end{array} \tag{IV.49}$$

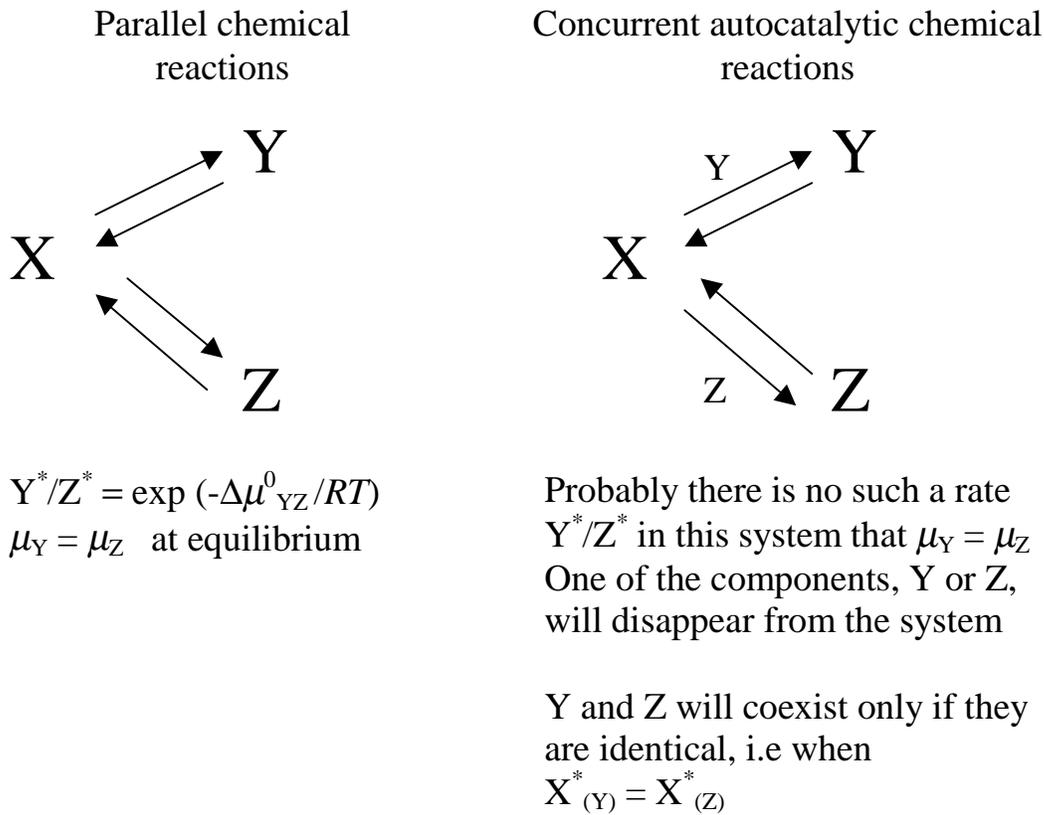

Parallel chemical reactions

$Y^*/Z^* = \exp(-\Delta\mu^0_{YZ}/RT)$
$\mu_Y = \mu_Z$ at equilibrium

Concurrent autocatalytic chemical reactions

Probably there is no such a rate $Y^*/Z^*$ in this system that $\mu_Y = \mu_Z$
One of the components, Y or Z, will disappear from the system

Y and Z will coexist only if they are identical, i.e when
$X^*_{(Y)} = X^*_{(Z)}$

*Figure 4*. A comparison of equilibrium states of two chemical processes – parallel and concurrent autocatalytic chemical reactions. In ideal conditions we conclude from equilibrium equations $\Delta\mu^0_{YX} = -RT \ln K_Y(Y^*)$ and $\Delta\mu^0_{ZX} = -RT \ln K_Z(Z^*)$ that $K_Y(Y^*)/K_Z(Z^*) = \exp(-\Delta\mu^0_{YZ}/RT)$; $\mu^0_i$ – the standard chemical potential of *i*th component. For parallel chemical reactions, which are shown on the left, $K_Y(Y^*)/K_Z(Z^*) \equiv K_Y/K_Z = Y^*/Z^*$. Autocatalytic parallel reactions coexist only if $K_Y(Y^*)/K_Z(Z^*) = 1$, i. e. when $\Delta\mu^0_{YZ} = 0$. The ratio $Y^*/Z^*$ at equilibrium may be any number depending on initial conditions – M, $Y_0$, $Z_0$. Anyway at a steady state we have $\mu_Y = \mu_Z$ due to the autocatalytic nature of chemical reactions.

These reactions may be presented by a system of differential equations

$dY/dt = b_1 X - d_1 Y$

$dZ/dt = b_2 X - d_2 Z$ (IV.50)

The reactions will reach equilibrium when $dY/dt = dZ/dt = 0$. This may be expressed in the terms of chemical potentials: at equilibrium chemical potentials of reacting components are equal. Indeed, the rate of a chemical reaction and chemical potentials of its components are closely related (Appendix 5). Near equilibrium we may write

$V = V^{\pm} A$ (IV.51)

where $V$ is the rate of chemical reaction; $V^{\pm}$ the rates of a forward or backward reaction at equilibrium; and $A = \mu_i - \mu_k$ is a difference of chemical potentials of reacting components.

The system shown in *Fig. 4a* reaches equilibrium after some time. Then chemical potentials of all components are equal, i.e. $\mu_X = \mu_Y = \mu_Z$. However, we cannot say the same if we deal with a system of concurrent chemical reactions (*Fig. 4b*). These reactions may be written in the form

$X + Y \rightarrow Y + Y;\ Y \rightarrow X$

$X + Z \rightarrow Z + Z;\ Z \rightarrow X$ (IV.52a)

Or we may write (IV.52a) as reversible reactions

$X \leftarrow\rightarrow Y$

$X \leftarrow\rightarrow Z$ (IV.52b)

if we denote $b_y = b_1 Y$ and $b_z = b_2 Z$. Differential equations for this system is

$dY/dt = b_1 XY - d_1 Y$

$dZ/dt = b_2 XZ - d_2 Z$ (IV.53)

Such a system will reach equilibrium only if $d_1/b_1 = d_2/b_2$. Indeed, we have

$Q_1 b_1 XY = K_1 d_1 Y;\qquad Q_1 = Y/X;\qquad K_1(Y) = (b_1/d_1)Y = Y/X^*_{(Y)}$

$Q_2 b_2 XZ = K_2 d_2 Z;\qquad Q_1 = Z/X;\qquad K_2(Z) = (b_2/d_2)Z = Z/X^*_{(Z)}$ (IV.54)

Here $K_1(Y)$ and $K_2(Z)$ are equilibrium constants of corresponding reactions which in this case really are functions that depend on arguments Y and Z. From (IV.54) we obtain that at equilibrium

$Q_i = K_i^* \Rightarrow Y^*/X^*_{(Y)} = (b_1/d_1)Y^*;\quad Z^*/X^*_{(Z)} = (b_2/d_2)Z^*$ (IV.55)

or

$X^*_{(Y)} = d_1/b_1;\quad X^*_{(Z)} = d_2/b_2$ (IV.56)

Because $M = X^* + Y^* + Z^* \Rightarrow M_{(Y)} = d_1/b_1 + Y^* + Z^*$ and $M_{(Z)} = d_2/b_2 + Y^* + Z^*$, then it is impossible to find such $Y^*$ and $Z^*$ that at the same time we would have $Y^* + Z^* = M - d_1/b_1$ and $Y^* + Z^* = M - d_2/b_2$ unless $d_1/b_1$ is equal to $d_2/b_2$.

If we return to the equation (IV.18)

$V = \varphi Y(1 - Y/\Phi)$ (IV.18)

we may rewrite it as

$V = \varphi \Phi^{-1} Y(\Phi - Y) \Rightarrow V = c\varphi \Phi^{-1} Y((M - c)/c - Y/c)$ (IV.57)

where $c$ is the equilibrium parameter in the expression $\Phi = M - c$. In (IV.57) $U/c$ is analogous to the equilibrium function $K$ (IV.54) of an autocatalytic chemical reaction. $\Phi/c$ in this case corresponds to $K^*$ value. For $n$ competing autocatalytic components $Y_i$ we have

$$V_i = c_{(i)}\varphi_i \Phi_i^{-1} Y_i ((M - c_{(i)})/c_{(i)} - \Sigma Y_k /c_{(i)}) \qquad (i = 1, \ldots, n) \tag{IV.58}$$

All components will coexist only if

$$(\Sigma Y_k^*)/c_{(i)} = (M - c_{(i)})/c_{(i)} = K_i^* \qquad (i = 1, \ldots, n) \tag{IV.59}$$

This can be attained if all $c_{(i)}$ are equal. It is clear that in this case directed competition, which is equal to $\Sigma \Phi_i^{-1} \Phi_k^{-1} (c_{(i)} - c_{(k)}) Y_i Y_k$ ($i = 1, \ldots, n$), does not reveal itself in the system.

So we may consider differences $c_{(i)} - c_{(k)}$ as some generalized forces and, thus, such forces are directed competition functions $\lambda_{ik}$.

What is relation between chemical potentials of competing components and directed competition? Take the ratio $d_Y/b_Y$. On the other hand $d_Y/b_Y = XdY/bXY$. From here we obtain that

$$ln(d_Y/b_Y) = ln X - \nu_Y \tag{IV.60}$$

For $\nu$ definition see Appendix 5. Rewrite (IV.60)

$$\nu_Y = ln X - ln(d_Y/b_Y) = ln X - ln X^*_{(Y)} = ln X - ln c_{(Y)} \tag{IV.61}$$

For another competitor Z we have

$$\nu_Z = ln X - ln c_{(Z)} \tag{IV.62}$$

Now substitute (IV.62) from (IV.61)

$$\nu_Y - \nu_Z = -(ln c_{(Y)} - ln c_{(Z)}) \tag{IV.63}$$

Because $\nu_Y - \nu_Z$ is proportional to $-(\mu_Y - \mu_Z)$ we may write

$$\mu_Y - \mu_Z = k (ln c_{(Y)} - ln c_{(Z)}) \tag{IV.64}$$

where $k$ is some coefficient.

Directed competition can be considered as a real force of biological evolution.

## V. Remarks

Now we return to LV competition equations (I.4)

$$dP_i/dt = r_i P_i (1 - \alpha_{i1} P_1 / K_i - \ldots - P_i / K_i - \ldots - \alpha_{in} P_n / K_i) \qquad (i = 1, \ldots, n) \tag{I.4}$$

As the logistic equation, this competition model may be written in the following form

$$dP_i/dt = r_i P_i - r_i \alpha_{i1} P_i P_1 / K_i - \ldots - r_i P_i^2 / K_i - \ldots - r_i \alpha_{in} P_i P_n / K_i \qquad (i = 1, \ldots, n) \qquad (V.1)$$

The term $r_i P_i$, as in the case of the logistic equation, describes the interaction of $P_i$ individuals with resources. As earlier, this term means that the amount of resources is infinite. Thus resources do not give any negative effect on the growth of population $P_i$. We merely may enhance or slow down the growth rate by maintaining the higher or lover amount of resources $R_0$ in the system according to the expression $r_i = b_i R_0 - d_i$. Claims that the logistic model or LV competition model describes competitive interactions between individuals for limited resources are false. LV competition model even does not require that all $P_i$ species consume the same food for which, therefore, they should compete. All species $P_i$ may use entirely different resources or they may consume the same food – the results will the same. This is because food is unlimited. Competitive interactions between individuals of the same or different species arise due to interference between these individuals. And only those negative interactions limit populations grow. This is an error of classical competition theory. The coefficient $K_i$ in the logistic or LV competition model does not have the meaning of carrying capacity of the species environment. These models miss the main component – the limited amount of resources, which determines how many individuals of a given species are able to survive. All other interactions between individuals of a given species and with individuals of other species modify this value increasing it or decreasing depending on the nature of interactions.

On the other hand we may take some initial model describing species interactions

$$dP_i/dt = r_i P_i - t_i P_i^2 - \Sigma_{k \neq i} s_{ik} P_i P_k \qquad (i, k = 1, \ldots, n) \qquad (V.2)$$

where $t_i$ and $s_{ik}$ show the intraspecific and interspecific interference interactions respectively. Transform this equation to the form (I.4) by writing

$$dP_i/dt = r_i P_i (1 - P_i/(r_i/t_i) - \Sigma_{k \neq i} (s_{ik}/t_i) P_k /(r_i/t_i)) \qquad (i, k = 1, \ldots, n) \qquad (V.3)$$

and further by denoting $K_i = r_i / t_i$ and $\alpha_{ik} = s_{ik} / t_i$. $K_i$ has a meaning only if $t_i \neq 0$. If $t_i = 0$, i.e. if individuals of species $P_i$ do not interfere, LV competition model has no sense.

Meanwhile equations (IV.34) have quite a different meaning. Functions $\Phi_i$ are really 'carrying capacities' according to their expression $\Phi_i = M - c_{(i)}$. They indeed show how many individuals of $i$th species may be supported in a given volume of environment due to relations with limited resources, including all other interactions described by a model.

At first consider system (V.2) without intraspecific interference interactions. If $t_i = 0$ ($i = 1, \ldots, n$) we have the equation

$$dP_i/dt = r_i P_i - \Sigma_{k \neq i} s_{ik} P_i P_k = r_i P_i (1 - \Sigma_{k \neq i} (P_k /(r_i/s_{ik}))) \qquad (i, k = 1, \ldots, n) \qquad (V.4)$$

This is not LV competition model. This conclusion follows from (V.3) expression if we assume that $t_i = 0$. In this case we should propose that $K_i = \infty$ what is inappropriate. Species which behavior is described by (V.4) equations will not coexist stably. Write equations (V.4) in the form

$$dP_i/dt = (b_i R_0 - d_i) P_i - \Sigma_{k \neq i} s_{ik} P_i P_k = b_i (M - \Sigma P_k) P_i - d_i P_i - \Sigma_{k \neq i} s_{ik} P_i P_k =$$

$$= (b_i M - d_i) P_i (1 - \Sigma P_k /(M - d_i/b_i) - \Sigma_{k \neq i} (s_{ik}/b_i) P_k /(M - d_i/b_i)) =$$

$$= \varphi_i P_i (1 - P_i / \Phi_i - \Sigma_{k \neq i} ((s_{ik} + b_i)/b_i) P_k / \Phi_i) \qquad (V.5)$$

We see that all parameters $\alpha_{ik} = (s_{ik} + b_i)/b_i > 1$. Therefore for every pair $\alpha_{ik}$ and $\alpha_{ki}$ we have $\alpha_{ik} \alpha_{ki} = ((s_{ik} + b_i)/b_i)((s_{ki} + b_k)/b_k) > 1$ (Appendix 6).

Thus species with only interspecific interference interactions cannot coexist stably. Contrary to this species with only intraspecific interference interactions will always coexist. This is clear from the following consideration. Let us take initial model, which is just another form of the logistic equation

$$dP_i/dt = r_iP_i - t_iP_i^2 \qquad (i, k = 1, \ldots, n) \tag{V.6}$$

(V.6) set of differential equations, not (I.4) model, describes the interaction of $n$ species growing on the same resource. All species will stably coexist in this system. Moreover they all will have the same equilibrium densities $K_i$ as if they would grow alone because resources are unlimited. If we want that competition for resources between species arises in (V.6) system we should deal with a system, which has the limited amount of total mass density M.

How equilibrium is reached in the logistic model may be seen from this example. We have

$$dP/dt = rP - tP^2 = bR_0P - dP - tP^2 = b(M - P)P - dP - tP^2 =$$

$$= (bM - d)P(1 - b^{-1}(b+t) P/(P+(R_0 - d/b))) \tag{V.7}$$

If $t_i > 0$ then the ratio $(b + t) / b$ is greater than 1. On the other hand the ratio $P/(P+(R_0 - d/b)) < 1$ increases when population P density grows. (We assume that the initial density of P is much less than the ratio $r/t$.) Thus there will be a moment in the evolution of this system when $((b + t) /b)$ $P/(P+(R_0 - d/b)) = 1$. Then logistic model (V.6) will reach a steady state (the density of P stops growing).

By the same chain of rearrangements as in (V.5) we obtain from (V.6)

$$dP_i/dt = (b_iR_0 - d_i) P_i - t_i P_i = (b_i M - d_i)P_i (1 - \Sigma P_k/(M - d_i/b_i) - (t_i/b_i)P_i /(M - d_i/b_i)) =$$

$$= \varphi_iP_i (1 - ((t_i + b_i)/b_i)P_i /\Phi_i - \Sigma_{k \neq i} P_k /\Phi_i) = \varphi_iP_i (1 - P_i /\tilde{\Phi}_i - \Sigma_{k \neq i} (b_i/(t_i + b_i))P_k /\tilde{\Phi}_i) \tag{V.8}$$

Here $\tilde{\Phi}_i = (b_i /(t_i + b_i))\Phi_i$. As $b_i /(t_i + b_i) < 1$ all species will survive in the system (Appendix 7).

Combining both intra- and interspecific interference interactions we may obtain various outcomes from our models. Take the model (V.2)

$$dP_i/dt = r_iP_i - t_iP_i^2 - \Sigma_{k \neq i} s_{ik}P_iP_k = (b_i M - d_i)P_i - b_i \Sigma P_k - t_i P_i^2 - \Sigma_{k \neq i} s_{ik}P_iP_k =$$

$$= \varphi_iP_i (1 - \Sigma P_k /\Phi_i - (t_i /b_i)P_i /\Phi_i - \Sigma_{k \neq i} (s_{ik}/ b_i) P_k /\Phi_i) =$$

$$= \varphi_iP_i (1 - P_i /\tilde{\Phi}_i - \Sigma_{k \neq i} ((s_{ik} + b_i)/( t_i + b_i)) P_k /\tilde{\Phi}_i) \tag{V.9}$$

where $\tilde{\Phi}_i = (b_i /(t_i + b_i))\Phi$. At least in the case when the product of all reciprocal interactions $\alpha_{ik}\alpha_{ki} = (s_{ik} + b_i)( t_i + b_i)^{-1}(s_{ki} + b_k)( t_k + b_k)^{-1}$ is less than 1 all species will coexist stably (Strobeck, 1973).

The examples analyzed above we wrote in such a form ((V.5), (V.8), (V.9)) that allows us to obtain a generalized expression of directed competition. Apply it to LV competition model (I.13)

$$dP_i/dt = r_iP_i (1 - \Sigma P_k/ K_k - \Sigma K_i^{-1}K_k^{-1} (\alpha_{ik}K_k - K_i) P_k) \qquad (i = 1, \ldots, n) \tag{I.13}$$

The term $\Sigma K_i^{-1}K_k^{-1} (\alpha_{ik}K_k - K_i)P_i P_k$ is generalized directed competition.

Now we may explain why in the case shown in *Fig. 5a* two species do not coexist and in the case shown in *Fig. 5b* they do. Simply in the second case generalized directed competition coefficients $\lambda_{ik} = K_i^{-1} K_k^{-1} (\alpha_{ik} K_k - K_i)$ for both species are less than zero. This means that species make a "positive" effect on each other by enhancing competitors growth rate. In the first case (Fig. 5a) the effect of generalized directed competition is positive for both species. Thus coexistence is impossible (the steady state is unstable).

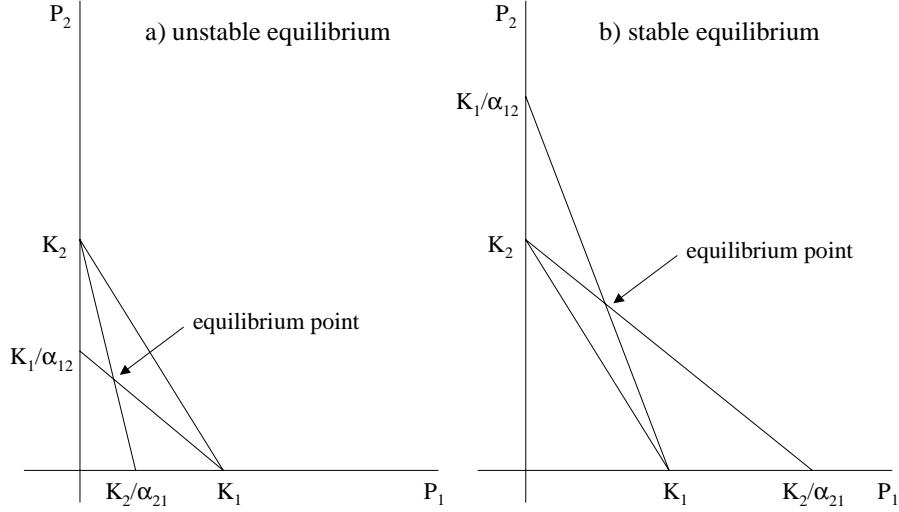

*Figure 5*. Two outcomes of two-species LV competition model.

Therefore the sign of directed competition determines which components in the system of competing species may be called winners or losers at any time. Again begin with the equation (IV.38) for d-competitors (it should be noted that we will obtain the same result if we consider c-competitors)

$$dY_i/dt = \varphi_i Y_i (1 - \Sigma Y_k/\Phi_k - \Sigma \lambda_{ik} Y_k) \tag{IV.38}$$

Rewrite it

$$(\varphi_i Y_i - dY_i/dt)/\varphi_i Y_i = \Sigma Y_k/\Phi_k + \Sigma \lambda_{ik} Y_k \tag{V.10}$$

The expression on the left is equivalent to the ratio $(X - X^*)/(M - X^*)$ and shows what part of available resources have been used already. Let us mark it $\Psi_i$. On the right side is the sum of indifferent and directed competition calculated for a density unit of $P_i$ species. We denote them by $\Psi$ and $\Psi_{i,\,dir}$ respectively. Thus we have

$$\Psi_i = \Psi + \Psi_{i,\,dir} \qquad (i = 1, \ldots, n) \tag{V.11}$$

If we multiply (V.11) equations by $Y_i$ and add them we get

$$\Psi_i Y_i = \Psi Y_i + \Psi_{i,\,dir} Y_i \qquad (i = 1, \ldots, n)$$

$$\Sigma \Psi_k Y_k = \Sigma \Psi Y_k + \Sigma \Psi_{k, dir} Y_k \qquad (V.12)$$

Because $\Sigma \Psi_{k, dir} Y_k = 0$ by identity, we obtain from (V.12) that

$$\Psi = (\Sigma \Psi_k Y_k)/(\Sigma Y_k) \qquad (V.13)$$

So $\Psi$ may be called specific $\Psi_k$ for the system of competing species. From here it follows that

$$\Psi_{i, dir} = \Psi_i - (\Sigma \Psi_k Y_k)/(\Sigma Y_k) = \Sigma(\Psi_i - \Psi_k)Y_k/\Sigma Y_k \qquad (V.14)$$

Thus $\Psi_{i, dir}$ is a mean deviation of all $\Psi_k$ ($k = 1, \ldots, n$) from a given $\Psi_i$. Therefore (V.11) may be rewritten as (Appendix 8)

$$\Psi_i = (\Sigma \Psi_k Y_k)/(\Sigma Y_k) + \Sigma(\Psi_i - \Psi_k)Y_k/\Sigma Y_k \qquad (i, k = 1, \ldots, n) \qquad (V.15)$$

Those competitors which have lover $\Psi_i$ ($\Psi_i < \Psi$) are favored in the system. $\Psi_i < \Psi$ means that $P_i$ species have more resources available for their populations growth. Species with negative $\Psi_{i, dir}$ are temporary winners in the system (see *Fig. 5b*). And species with positive $\Psi_{i, dir}$ suffer a loss from competition with other species (see *Fig. 5b*). The system of competing species with both c- and d-competitors is in equilibrium only when $\Psi_i = \Psi$ for all $i = 1, \ldots, n$. Or more generally, if all species in a system coexist permanently then

$$\lim_{t \to \infty} t^{-1} \int_0^t \Psi_i \, d\tau = \lim_{t \to \infty} t^{-1} \int_0^t \Psi \, d\tau \qquad (V.16)$$

Now, at last, we have arrived at a point where obvious and simple conclusions can be made. Return to the equation (V.11)

$$\Psi_i = \Psi + \Psi_{i, dir} \qquad (i = 1, \ldots, n) \qquad (V.11)$$

At first let us assume that our system of c- and d-competitors is a "normal" system. This means that the total mass flow is from resources to consumers, i.e. left to right in the equation $X \leftarrow\!\!\!\rightarrow Y$. By accepting this agreement we simply exclude the situation where the system is overloaded, i.e. where the concentration of at least some of $Y_i$'s is more that the system can sustain.

I state here that contemporaneous evolutionary thoughts, which have their roots in the works of Darwin and Malthus, rest on the $\Psi$ value defined according to the equation (V.11). Indeed, if we want for natural selection to operate, the system should reach a state where resources become limited and thus competition between individuals begin. This state of the system is measured with the function $\Psi$. And, strictly speaking, competition should occur between individuals when $\Psi = 1$. But in the "normal", not disturbed, system this may occur only if all competitors coexist. In this case either they are identical or there are left only one c-competitor and one d-competitor in the system. In any case c-competitors and d-competitors would be identical and selection has nothing to do with them.

Contrary to what scientists believe the principle of natural selection has no background to be a scientific model of biological evolution. Natural selection theory is simply some kind of an anthropomorphic view of the world. Actually it does not follow from the Darwin's argument that all living things have an unlimited potential for growth. From this someone could conclude that species should struggle for their survival. On the other hand I may state that from the same proposition the conclusion that species have no reason to struggle follows (because potential unlimited growth means that here is nothing in the system which can suppress species).

It seems that the process of natural selection can be presented with a scheme given in *Fig. 4*. The first impression is such that the fittest species, i. e. a species, which has a higher $\Phi_i$ value, is the winner. But this is very superficial impression. Here I give another explanation for this competitive exclusion event. The result of competition is not the survival of the fittest but the disappearance of differences between competitors, Y and Z. It may be done by two ways. First, one of the species with a lower $\Phi_i$ value must extinct. Then surviving competitor would become identical because individuals of the same species are considered to be identical. Second, Both species, Y and Z, may coexist if they are (or become) identical.

One more remark. Natural selection idea arose from the so called artificial selection process. If natural selection does not exist how we can explain artificial selection? We can do that in a simple manner. Man does not select individuals. He just changes the birth and death rates of selected populations, therefore changing their $\Phi_i$ values.

My conclusion is that only functions $\Psi_{i, dir}$ are responsible for the Darwinian evolution of competing species. Thus, the result of evolution is not better-adapted individuals, but a system with identical components. We have a clear expression of species fitness – their $\Phi_i$ (or $c_{(i)}$) values. Of course species with a higher $\Phi_i$ (or lower $c_{(i)}$) wins. But evolution may lead to the highest values of fitness if only one species (or perhaps one super-individual) would exist on earth. However, this is quite impossible because the degradation of individuals is physically inevitable. So the emergence of new species is an ordinary process of moving toward a less ordered state. When different species present in the system we should pay attention to the following things regarding their interactions. At first we should remember that c- and d-competitors evolve in different directions. d-competitors increase their $\Phi_i$ values while c-competitors decrease them. Second, in general, interactions between species decrease each other fitness. The individual interacting components – species – cannot evolve toward some extreme fitness value. But this loss in fitness leads to the increases of species number. Probably the huge biodiversity indeed maintains the whole system at a relatively equilibrium state (Coste *et al*., 1978; Anthony and Stuart, 1983; Rooney *et al*., 2006; Ives and Carpenter, 2007). The more potential interactions exist in the system the easier species can control their behavior so that to make themselves approximately identical.

The principle of natural selection cannot be applied to chemical world; so it cannot explain the origin of life. Therefore a breach exists between chemical and biological worlds. Nevertheless we may repair this proclaiming that natural selection principle is a universal law. Then selection may occur between chemical molecules (Eigen and Schuster, 1979). However, in this paper another view on biological evolution and the maintaining of species diversity is presented. I reject not only the "cosmological" nature of natural selection principle but the principle itself. The initial steps in the emergence of life probably were some parallel autocatalytic chemical reactions. The crucial event for biological evolution to begin was the emergence of self-replicating chemical entities (Eigen and Schuster, 1979; Diener, 1989; Woese, 1998). As soon as the simplest self-replicating systems arise it is quite possible that their complexity will increase. The more complex self-replicating units are the more "degrees of freedom" they have to manipulate their behavior. This enhances the possibility to maintain competitors similarity. Therefore different forms have more chances to survive.

## Appendices

### Appendix 1

Model of self-replication (IV.18) not necessary describes the transfer of mass between two components. Suppose we have some population of mass units, which may exchange of some "information" or "pathogen". If we denote the mass density of this population by Y, "clear" individuals by $Y_0$, "infected" individuals by $Y_1$, we may write

$Y_0 + Y_1 \rightarrow Y_1 + Y_1$

$Y_1 \rightarrow Y_0$ (A1.1)

$dY_1/dt = bY_0 Y_1 - d Y_1$ (A1.2)

where $d$ is the rate by which every single unit "forget" the message transferred to him or recover from the infection. From (A1.2) we obtain

$dY_1/dt = \varphi Y_1(1 - Y_1/\Phi)$ (A1.3)

Here $\varphi = bY - d$, $\Phi = \varphi/b = Y - d/b$. The form $Y_1$ of component Y will be sustained if $d/b < Y$. In the case of a "pathogen" we would like to increase $d$ and decrease $b$. If we deal with some important content of "information" we will strive to decrease $d$ and increase $b$.

### Appendix 2

When biologists deal with LV predator-prey model

$dX/dt = rX - fXY$

$dY/dt = bXY - dY$ (A2.1)

they assume that $f \neq b$ (usually $f > b$). This is explained as the loss of energy when transferred from prey to predator. I avoid such an inappropriate formulation of the model. From (A2.1) it is impossible to measure the true value of $c$ in the expression $\Phi = M - c$ (and, thus, $\Phi$ value itself). To obtain this value we should reformulate (A2.1) in the form

$dX/dt = rX - fXY$

$dY/dt = fXY - ((f - b)X - d)Y$ (A2.2)

Of course the behavior of (A2.1) and (A2.2) systems are identical. But the function $b$ in the model (A2.1) contains the processes (respiration, which is equal to $(f - b)XY$) which should not be included in it when we derive the equations of (IV.18) type.

**Appendix 3**

Function

$$h = r(e^{\eta} - \eta - 1) + d(e^{\xi} - \xi - 1) = const \tag{A3.1}$$

models the periodic behavior of the LV predator-prey system

$$dS/dt = rS - gSP$$

$$dP/dt = bSP - dP \tag{A3.2}$$

Here $\eta = \ln P - \ln P^*$ and $\xi = \ln S - \ln S^*$, where $S^* = d/b$ and $P^* = r/g$ are equilibrium densities. When oscillations occur near the stationary point $(S^*, P^*)$ $h$ obtains the following form

$$h = 2^{-1}(r\eta^2 + d\xi^2) = const \tag{A3.3}$$

(A3.3) describes the trajectory of a system with two degrees of freedom, which corresponds to the model

$$d^2\xi/dt^2 = -rd\xi$$

$$d^2\eta/dt^2 = -rd\eta \tag{A3.4}$$

**Appendix 4**

Here we look at another way of obtaining an equation of the LV competition model type. This equation is used by Eigen and Schuster (1979). Let us consider $n$ components $Y_1, \ldots, Y_n$ growing in a turbidostat according to the equation

$$dY_i/dt = r_iY_i = b_iY_i - d_iY_i \tag{A4.1}$$

Here $d_i = d$ for all index $i$. According to the turbidostat definition $dY_i = dV/(T\,Y_i dt)$, where $dV/dt$ is the culture remove rate from the system, and $T = const$ is a total volume of culture. Thus

$$dY_i/dt = r_iY_i = b_iY_i - dV/(T\,Y_i dt) \tag{A4.2}$$

The total concentration of all components $K$ is held constant. Then $T\Sigma\,b_kY_k = \Sigma Y_k(dV/dt)$. From here we find that $dV/dt = (T\Sigma\,b_kY_k)/\Sigma Y_k$. So

$$dY_i/dt = b_iY_i - ((\Sigma\,b_kY_k)/\Sigma Y_k)Y_i \tag{A4.3}$$

and

$$dY_i/dt = b_iY_i(1 - (\Sigma(b_k/b_i)\,Y_k)/K) \tag{A4.4}$$

The component, which has the highest $b_i$, wins in the contest. This is the reason why Malthusian parameter – the growth rate $b_i$ (or more generally $r_i$) – is considered as a fitness parameter (Biebricher, Eigen, 2005). This is an error. The fitness of $Y_i$ component is the ratio $d_i/b_i$, not the Malthusian growth rate parameter. Indeed, from (A4.3) we obtain

$dY_i/dt = b_i R\, Y_i/R - ((\Sigma\, b_k Y_k) / \Sigma Y_k)Y_i = b_i/R(M - \Sigma Y_k)Y_i - ((\Sigma\, b_k Y_k) / \Sigma Y_k)Y_i =$

$= (b_i\, R^{-1}M - (\Sigma\, b_k Y_k) / \Sigma Y_k)Y_i\, (1 - \Sigma Y_k /(M - (R\Sigma\, b_k Y_k)/( b_i \Sigma Y_k))$ (A4.5)

Here $\Phi_i = M - c_{(i)}$, where $c_{(i)} = (R\Sigma\, b_k Y_k)/( b_i\, \Sigma Y_k)$. $c_{(i)}$'s depend only on one parameter − $b_i$: $c_{(i)}$'s decreases when $b_i$ increases. This creates a false impression that the Malthusian growth rate describes the component fitness.

**Appendix 5**

Consider a formal reversible chemical reaction

$X \leftarrow\rightarrow Y$ (A5.1)

We may think about reaction $mX \leftarrow\rightarrow nY$ but it is not necessary. The result will be the same.
   The model of reaction (A5.1) is

$dX/dt = -k_1 X + k_2 Y$

$dY/dt = k_1 X - k_2 Y$ (A5.2)

Denote $V^+ = k_1 X$ and $V^- = k_2 Y$. If we denote $K = k_1/k_2 = Y^*/X^*$ and $Q = Y/X$, then we obtain $QV^+ = KV^-$. The driving force for this reaction is $A = \mu_X - \mu_Y$ (Caplain, Essig, 1983). Thus we have ($\mu_X^*$ is chemical potential at equilibrium)

$A = (\mu_X - \mu_X^*) - (\mu_Y - \mu_Y^*) = RT\, ln(X/X^*) - RT\, ln(Y/Y^*) = RT\, ln(XY^*/X^*Y) = RT\, ln(K/Q)$ (A5.3)

or

$A = RT\, ln(V^+/V^-)$ (A5.4)

Let us say that $V^+ = exp(v^+)$ and $V^- = exp(v^-)$. Then

$A = RT\, ln(exp(v^+) / exp(v^-)) = RT\, (v^+ - v^-)$ (A5.5)

So we have

$V = (V/A)\, A = R^{-1}T^{-1}((V^+ - V^-)/(v^+ - v^-))A = \omega\, R^{-1}T^{-1}((exp(v^+) - exp(v^-)) / (v^+ - v^-))A$ (A7.6)

where $\omega$ is a unit rate constant. As

$exp(v^+) - exp(v^-) = chv^+ + shv^+ - chv^- - shv^- = (shv^+ - shv^-) + (chv^+ - chv^-) =$

$= 2sh((v^+ - v^-)/2)\, ch((v^+ + v^-)/2) + 2sh((v^+ + v^-)/2)\, sh((v^+ - v^-)/2) =$

$= 2sh((v^+ - v^-)/2)\, (ch((v^+ + v^-)/2) + sh((v^+ + v^-)/2)) = 2sh((v^+ - v^-)/2)\, exp((v^+ + v^-)/2) =$

$$= 2sh((v^+ - v^-)/2)\ exp(v^+/2)\ exp(v^-/2)$$

therefore by denoting $v^+ - v^- = v$ we obtain

$$V = R^{-1}T^{-1}\ (V^+V^-)^{1/2}\ (sh(v/2)/(v/2))\ A \tag{A5.7}$$

Because $de^x/dx = e^x$ the expression $(V^+V^-)^{1/2}\ (sh(v/2)/(v/2))$ means some hypothetical equilibrium state of a reaction. Near true equilibrium we may write

$$V = R^{-1}T^{-1}\ (V^+V^-)^{1/2}\ A \tag{A5.8}$$

or even

$$V = (RT)^{-1}V^{\pm}A \tag{A5.9}$$

Here $V^{\pm} = V^+ = V^-$ is the rate of forward and backward processes at equilibrium.
     Further, we might assume that the function $(V^+V^-)^{1/2}\ (sh(v/2)/(v/2))$ in (A5.7) may be expressed as a weighted geometric mean of $V^+$ and $V^-$. Indeed, we write

$$exp(v^+/2)\ exp(v^-/2)\ (sh(v/2)/(v/2)) = (exp(v^+) - exp(v^-))/(v^+ - v^-) = exp(v^+)^{r^+}\ exp(v^-)^{r^-} \tag{A5.10}$$

Here $r^+ = s^+/(s^+ + s^-)$ and $r^- = s^-/(s^+ + s^-)$. From here

$$r^- = (v^+ - (ln(exp(v^+) - exp(v^-))/(v^+ - v^-)))/(v^+ - v^-) \tag{A5.11}$$

By the same way we obtain that

$$r^+ = ((ln(exp(v^+) - exp(v^-))/(v^+ - v^-)) - v^-)/(v^+ - v^-) \tag{A5.12}$$

**Appendix 6**

(V.4) equations we write for two interacting species

$$dP_1/dt = r_1 P_1 - s_{12}P_1 P_2$$

$$dP_2/dt = r_2 P_2 - s_{21}P_2 P_1 \tag{A6.1}$$

We have

$$(\partial(dP_1/dt)/\partial P_1)_{eq} = (r_1 - s_{12} P_2)_{eq} = 0$$

$$(\partial(dP_1/dt)/\partial P_2)_{eq} = (-s_{12} P_1)_{eq} = -(s_{12}/s_{21})r_2$$

$$(\partial(dP_2/dt)/\partial P_1)_{eq} = (-s_{21} P_2)_{eq} = -(s_{21}/s_{12})r_1$$

$$(\partial(dP_2/dt)/\partial P_2)_{eq} = (r_2 - s_{21} P_1)_{eq} = 0$$

Thus $\alpha^2 - r_1 r_2 = 0$. From here $\alpha_{1,2} = \pm (r_1 r_2)^{1/2}$. Equilibrium of (A6.1) system is an unstable saddle point.

**Appendix 7**

From (V.6) equations we construct two species competition system

$$dP_1/dt = r_1 P_1 - t_1 P_1^2$$

$$dP_2/dt = r_2 P_2 - t_2 P_2^2 \qquad (A7.1)$$

Here

$$(\partial(dP_1/dt)/\partial P_1)_{eq} = (r_1 - 2t_1 P_1)_{eq} = -r_1$$

$$(\partial(dP_1/dt)/\partial P_2)_{eq} = (-s_{12} P_1)_{eq} = 0$$

$$(\partial(dP_2/dt)/\partial P_1)_{eq} = (-s_{21} P_2)_{eq} = 0$$

$$(\partial(dP_2/dt)/\partial P_2)_{eq} = (r_2 - s_{21} P_1)_{eq} = -r_2$$

Then $(r_1 + \alpha)(r_2 + \alpha) = 0$ or $\alpha^2 + (r_1 + r_2)\alpha + r_1 r_2 = 0$. So $\alpha_1 = -r_2$ and $\alpha_2 = -r_1$. Because $r_1, r_2 > 0$ the equilibrium point is stable.

**Appendix 8**

We may begin with some relative measure $\Psi_i$ describing how much resources are depleted. We have

$$\Psi_i = \Psi_i \Sigma U_k / \Sigma U_k = \Sigma U_k /(\Sigma U_k/\Psi_i) \qquad (A8.1)$$

If the system reaches a stable state then $\Psi_i = 1$ and $\Sigma U_k = \Sigma U_k/\Psi_i$. Thus the expression $\Sigma U_k/\Psi_i$ is some maximum density of all populations growing on a given amount of resources. Denote it by $\Phi_i$. Now let us write the following equation

$$\Psi_i = (\Sigma \Psi_k U_k) / \Sigma U_k + (\Sigma(\Psi_i - \Psi_k)U_k)/(\Sigma U_k) \qquad (A8.2)$$

or, substituting $\Sigma U_k/\Psi_l$ with $\Phi_i$,

$$\Psi_i = \Sigma U_k/\Phi_k + \Sigma((\Phi_k - \Phi_i)/\Phi_i\Phi_k)U_k \qquad (A8.3)$$

If we assume that functions $\Psi_i$ may be evaluated by measuring the relative growth rates of corresponding populations, we may write the expressions for $\Psi_i$

$$\Psi_i = (V_{max} - V_t)/V_{max} \qquad (A8.4)$$

thus obtaining the usual rate equation

$$V_t = V_{max}(1 - \Sigma U_k/\Phi_k - \Sigma((\Phi_k - \Phi_i)/\Phi_i\Phi_k)U_k) \qquad (A8.6)$$